\documentclass[cameraready]{Interspeech}

\title{VoxEffects: A Speech-Oriented Audio Effects Dataset and Benchmark}

\author[affiliation={}, orcid=0000-0003-1003-6093, equalcontribution]{Zhe}{Zhang}
\author[affiliation={}, orcid=0000-0003-2235-8655, equalcontribution]{Yigitcan}{Özer}
\author[affiliation={}, orcid=0000-0003-2752-3955]{Junichi}{Yamagishi}

\address{
    National Institute of Informatics, Tokyo, Japan
}

\email{\{zhe, yiitozer, jyamagis\}@nii.ac.jp}

\keywords{audio effects, speech processing, speech dataset}

\usepackage{comment}
\usepackage{multirow}
\usepackage{xcolor}
\usepackage[normalem]{ulem}
\usepackage{url}

\begin{document}

\maketitle

\begin{abstract}
Speech audio in the wild is often processed by post-production effects, but existing speech datasets rarely provide precise annotations of effects and parameters, limiting systematic study. We introduce VoxEffects, a speech audio effects dataset that pairs produced speech with exact effect-chain supervision at multiple granularities. VoxEffects supports speech-oriented audio effect identification: given a produced waveform, infer which effects are present and how they are applied. Built from minimally edited clean speech, it provides an extensible rendering pipeline for both offline synthesis and on-the-fly rendering for efficient training and evaluation. The audio effect identification benchmark includes effect presence detection, preset classification, and intensity prediction, with a robustness protocol covering capture-side and platform-side degradations. We provide an AudioMAE-based multi-task baseline and analyses of domain shift, robustness, input duration, and gender fairness.
\end{abstract}

\section{Introduction}
\label{sec:intro}

Real-world speech audio is rarely ``raw.'' 
In practice, recordings are typically processed by a small set of quality-oriented post-production effects.
These operations improve intelligibility and perceived broadcast quality, but also introduce audio artifacts and shift signal statistics that downstream systems rely on.

In this paper, we study \textbf{audio effect identification (AEI)} for speech: given a processed waveform, infer \emph{which} effects are present and \emph{how} they are applied.
AEI supports production-aware content understanding, audio engineering assistance, educational tools for ear training, and audio forensics where production can confound attribution and authenticity judgments.

Despite these motivations, AEI remains underexplored as a standardized speech task with post-processing supervision.
Audio forensics has largely focused on integrity cues and binary ``real vs.\ fake'' decisions rather than attributing realistic post-production cues~\cite{MuellerEtAl24_VoiceAuthenticity_Interspeech, BevinamaradShirldonkar20_AudioForgeryReview_ICOEI, ZakariahEtAl18_AudioForensicsSurvey_MTA, EsquefEtAl14_EditDetection_ENF_TIFS}.
Recent perspectives call for richer taxonomies of voice edits beyond binary judgments~\cite{MuellerEtAl24_VoiceAuthenticity_Interspeech}.
Our study is complementary to these directions: instead of detecting manipulation, it attributes common \emph{benign} processing operations that routinely appear in distribution pipelines, a practically pervasive but underexplored problem.

Parallel research focuses on differentiable or learnable audio effect modeling and blind parameter estimation~\cite{ComunitaSteinmetzReiss25_DifferentiableAFX_Frontiers, PeladeauPeeters24_BlindEffectsEstimation_ICASSP, GoetzEtAl24_BlindAcousticEstimation_IWAENC}.
Music-oriented research has examined effect recognition and parameter estimation for instruments such as guitar~\cite{GuoMcFee23_CascadedGuitarEffects_DAFx, ComunitaSR21_GuitarEffects_JAES, JuergensEtAl20_GuitarEffects_DAFx, AbesserEtAl10_GuitarAudioEffects_AES} and singing vocal~\cite{YuEtAl25_DiffVox_DAFx}.
Other studies consider removing unknown effects~\cite{RiceSFD23_RemFX_WASPAA} or predicting effect parameters from language descriptions~\cite{DohEtAl25_LLM2Fx_WASPAA}.
However, these efforts typically target music production regimes, specific implementations, or do not evaluate robustness to distribution artifacts such as resampling and lossy compression that are central in speech pipelines.

\begin{figure}[t]
    \centering
    \includegraphics[width=\columnwidth]{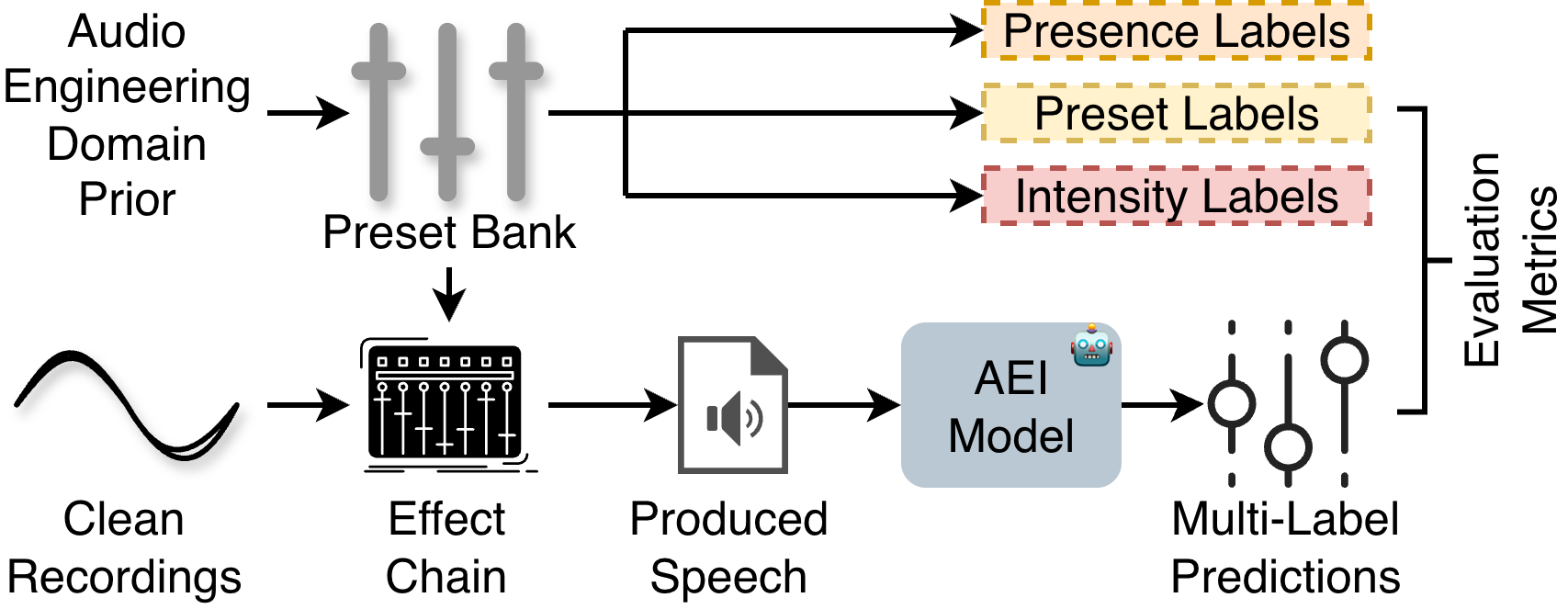}
    \caption{Overview of VoxEffects framework: dataset creation via a speech post-processing chain with curated presets, and benchmark evaluation via multi-granularity prediction.}
    \label{fig:overview}
\end{figure}

To address this gap, we introduce \textbf{VoxEffects}, a speech-oriented audio effects dataset and benchmark built from clean speech recordings and a post-production chain with curated preset banks, derived from domain knowledge in speech production and audio engineering, and evaluated under realistic capture- and platform-side degradations.
Fig.~\ref{fig:overview} illustrates the VoxEffects framework, which includes a reproducible renderer that supports both offline dataset release and scalable on-the-fly generation, enabling multi-granularity supervision (presence, preset, and intensity) under controlled robustness protocols.
VoxEffects and our benchmark protocols are released\footnote{\url{https://github.com/nii-yamagishilab/VoxEffects}}.

Our main contributions include the following:
\begin{itemize}
    \item \textbf{VoxEffects dataset and audio effect renderer.}
    A quality-oriented speech post-production chain with curated presets and a reproducible rendering pipeline for multi-granularity supervision under controlled train/test degradation settings.
    \item \textbf{Speech AEI benchmark with multiple tasks.}
    Multi-label effect presence detection, preset classification, number-of-effects counting prediction, and intensity regression.
    \item \textbf{Baseline and analysis.}
    An AudioMAE-Fx model based on AudioMAE~\cite{HuangEtAl22_AudioMAE_NeurIPS}, with analysis of cross-corpus generalization, robustness to capture- and platform-side degradations, effect-wise performance, input duration, and gender fairness.
\end{itemize}

\section{VoxEffects Dataset}
\label{sec:dataset}
VoxEffects is a speech-oriented audio effects dataset and benchmark that pairs produced speech with multi-granularity supervision.
It includes a reproducible pipeline supporting both offline synthesis and on-the-fly rendering for scalable training and evaluation under multiple controlled robustness settings.

\subsection{Design priors and effect chain}
\label{subsec:data_priors}
VoxEffects is designed around prior knowledge of speech post-production processing in audio engineering.
First, practical speech pipelines typically follow a fixed or near-fixed effect ordering, where upstream stages influence downstream cues~\cite{Mysore15_ProductionQualitySpeech_SPL,izhaki_mixing_audio_4ed}.
Second, speech production usually relies on a small set of engineer-tuned \emph{presets} that apply quality-oriented processing, rather than the diverse creative effects and settings common in music production~\cite{izhaki_mixing_audio_4ed,senior_mixing_secrets,katz_mastering_audio}.

Motivated by these observations, we model a canonical speech post-production chain with six effects:
\textbf{\textit{DN}} (denoising) $\rightarrow$
\textbf{\textit{DRC}} (dynamic range compression) $\rightarrow$
\textbf{\textit{EQ}} (equalization) $\rightarrow$
\textbf{\textit{DS}} (de-essing) $\rightarrow$
\textbf{\textit{RVB}} (reverberation) $\rightarrow$
\textbf{\textit{LIM}} (limiting),
and denote the effect set by
$\mathcal{V}=\{\textbf{\textit{DN}},\textbf{\textit{DRC}},\textbf{\textit{EQ}},\textbf{\textit{DS}},\textbf{\textit{RVB}},\textbf{\textit{LIM}}\}$.
For each effect $e\in\mathcal{V}$, we define a discrete preset set
$\mathcal{P}_e=\{p_{e,0},\ldots,p_{e,K_e}\}$, where $p_{e,0}$ denotes \texttt{bypass}, $K_e$ is the number of non-bypass presets, and each $p_{e,k}$ corresponds to a speech-oriented parameter vector for effect $e$.

For example, $\mathcal{P}_{DN}$ includes \texttt{bypass} ($p_{\text{DN},0}$) and two noise-gate presets as follows:
\begin{itemize}
\item[] {\footnotesize\ttfamily
\begin{tabular}{@{}l@{}l@{}}
$p_{\text{DN},1}$\;=\; &
NoiseGate(threshold=-50\,dB, ratio=4,\\
& \phantom{NoiseGate(}attack=2\,ms, release=50\,ms),\\
$p_{\text{DN},2}$\;=\; &
NoiseGate(threshold=-40\,dB, ratio=8,\\
& \phantom{NoiseGate(}attack=1\,ms, release=100\,ms).
\end{tabular}
}
\end{itemize}

A complete configuration is represented by the preset tuple
\begin{equation}
\mathbf{p}=(p_{\text{DN}}, p_{\text{DRC}}, p_{\text{EQ}}, p_{\text{DS}}, p_{\text{RVB}}, p_{\text{LIM}}).
\end{equation}
We implement the audio effect renderer $\mathcal{R}(\cdot;\mathbf{p})$ using the \texttt{Pedalboard}~\cite{sobot_pedalboard_2023} audio effects library.

\subsection{Preset bank and combinatorial coverage}
\label{subsec:preset_bank}

We instantiate $\{\mathcal{P}_e\}_{e\in\mathcal{V}}$ as a compact preset bank per effect:
\textbf{\textit{DN}}: 3,
\textbf{\textit{DRC}}: 5,
\textbf{\textit{EQ}}: 7,
\textbf{\textit{DS}}: 3,
\textbf{\textit{RVB}}: 4,
\textbf{\textit{LIM}}: 2,
yielding \textbf{2520} preset combinations.
The bank is designed to balance coverage and tractability for speech AEI. Preset counts reflect how diversely each effect is used in speech post-production (e.g., more \textbf{\textit{EQ}} and \textbf{\textit{DRC}} variants than \textbf{\textit{LIM}}). The presets span mild-to-moderate, quality-oriented regimes rather than aggressive creative settings. Keeping the bank sizes moderate prevents a combinatorial explosion, making the 2520-way preset classification challenging but feasible.
These presets are curated to reflect common \emph{benign} operating regimes used to improve clarity and loudness consistency in distributed speech (e.g., light vs.\ moderate denoising; gentle vs.\ broadcast-style compression; de-essing with mild vs.\ stronger control).

\subsection{Distribution degradations for robustness}
\label{subsec:dist_degrad}

To reflect real deployment conditions, VoxEffects includes an optional degradation module $\mathcal{D}(\cdot)$ that simulates common capture and platform artifacts (e.g., additive background noise, resampling ladders, and lossy codecs).
Degradations are applied at two stages relative to effect rendering: \emph{capture-side} (before the effect chain) and \emph{platform-side} (after the effect chain).
Given a clean speech waveform $\mathbf{x}\in\mathbb{R}^{T}$ and a preset tuple $\mathbf{p}$, we define
\begin{equation}
\tilde{\mathbf{x}}=\mathcal{D}_{\text{post}}\left(\mathcal{R}(\mathcal{D}_{\text{pre}}(\mathbf{x});\mathbf{p})\right),
\end{equation}
where $\mathcal{D}_{\text{pre}}$ and $\mathcal{D}_{\text{post}}$ are identity (no degradation) or $\mathcal{D}$.
This yields five reproducible settings used throughout the benchmark:
\textbf{None} (no degradation),
\textbf{Pre-only} ($\mathcal{D}_{\text{pre}}{=}\mathcal{D},\,\mathcal{D}_{\text{post}}{=}\mathrm{Id}$),
\textbf{Post-only} ($\mathcal{D}_{\text{pre}}{=}\mathrm{Id},\,\mathcal{D}_{\text{post}}{=}\mathcal{D}$),
\textbf{Either} ($(\mathcal{D}_{\text{pre}},\mathcal{D}_{\text{post}})$ is sampled independently per example as
$(\mathcal{D},\mathrm{Id})$ or $(\mathrm{Id},\mathcal{D})$ with probability $1/2$ each),
and \textbf{Both} ($\mathcal{D}_{\text{pre}}{=}\mathcal{D},\,\mathcal{D}_{\text{post}}{=}\mathcal{D}$).

\subsection{Source audio and splits}
\label{subsec:source_splits}
VoxEffects is created by processing raw speech recordings with minimal room coloration or prior production edits.
To ensure such acoustic conditions, we build VoxEffects from three source corpora: DAPS~\cite{Mysore15_ProductionQualitySpeech_SPL}, EARS~\cite{RichterEtAl24_EARS_Interspeech}, and TSP~\cite{Kabal18_TSP}, which are recorded under anechoic or near-anechoic conditions.
Starting from clean recordings reduces confounding effects of unknown upstream processing and makes VoxEffects naturally extensible: new raw corpora and expanded preset banks can be added without changing the dataset interface.

For each source corpus, we split the utterances into train/validation/test with an \textbf{8:1:1} ratio.
We report \textbf{in-domain (ID)} performance by evaluating on the test splits of these three corpora, and \textbf{out-of-domain (OOD)} generalization performance using VCTK\,\cite{VeauxYM19_VCTK} as a separate test-only corpus rendered under the same chain, presets, and degradation settings.

\section{Audio Effect Identification Benchmark}
\label{sec:task}

\subsection{Audio effect identification tasks}
\label{subsec:tasks}

\smallskip\noindent\textbf{Main task: effect presence detection.}
We define a binary presence vector $\mathbf{y}\in\{0,1\}^{|\mathcal{V}|}$ from the effect-chain configuration:
$y_e{=}1$ if effect $e$ is active (i.e., its preset is not \texttt{bypass}), and $y_e{=}0$ otherwise.
Given an input waveform $x$, the model predicts per-effect probabilities $\hat{\mathbf{y}}\in[0,1]^{|\mathcal{V}|}$.

\smallskip\noindent\textbf{Sub-task 1: fine-grained preset classification.}
We treat each preset tuple $\mathbf{p}$ as one class among
$C=\prod_{e\in\mathcal{V}}|\mathcal{P}_e|$ configurations (here $C=2520$),
and report Top-1/Top-5 accuracy.

\smallskip\noindent\textbf{Sub-task 2: counting the number of active effects.}
We predict the number of active effects $n=\sum_e y_e \in \{0,\dots,|\mathcal{V}|\}$.

\smallskip\noindent\textbf{Sub-task 3--4: intensity regression.}
We map each preset to a normalized strength via \mbox{$\alpha_e:\mathcal{P}_e\rightarrow[0,1]$}
(e.g., \mbox{$\alpha_e(p_{e,k})=k/K_e$ and $\alpha_e(p_{e,0})=0$)}.
This yields (i) a scalar overall intensity defined as the mean normalized preset index for all effects, and
(ii) a per-effect intensity vector $\mathbf{t}\in[0,1]^{|\mathcal{V}|}$ with $t_e=\alpha_e(p_e)$.
We evaluate models on both scalar and vector intensity prediction.

\subsection{Evaluation metrics}
\label{subsec:metrics}

\smallskip\noindent\textbf{Effect presence.}
We report macro-averaged accuracy across effects, denoted as Acc$_\text{macro}$, per-effect accuracy, and the example-level exact match ratio (EMR). Unless stated otherwise, probabilities are thresholded at 0.5.

\smallskip\noindent\textbf{Preset classification.}
We report Top-1 and Top-5 accuracy values for the fine-grained present classification task.

\smallskip\noindent\textbf{Number of active effects.}
We report classification accuracy.

\smallskip\noindent\textbf{Effect intensity regression.}
For scalar intensity (Sub-task~3), we report the mean absolute error MAE$_\text{overall}$.
For vector intensity (Sub-task~4), we report the mean per-effect MAE, denoted MAE$_\text{mean}$, and optionally per-effect MAE.

\smallskip\noindent\textbf{Domain shift and robustness.}
We report the results under (i) cross-corpus generalization (ID $\rightarrow$ OOD) and
(ii) robustness to distribution artifacts via controlled degradations $\mathcal{D}(\cdot)$ applied before and/or after rendering.

\begin{table*}[htbp]
\centering
\caption{Benchmarking results under different train/test augmentations. Each cell reports In-Domain / Out-of-Domain performance.}
\label{tab:robustness_idood_2col}
\begin{tabular}{llccccccc}
\toprule
\multicolumn{1}{l}{Test} & \multicolumn{1}{l}{Train}
& \multicolumn{2}{c}{Effect Presence}
& \multicolumn{2}{c}{Effect Preset}
& \multicolumn{1}{c}{\#Active}
& \multicolumn{2}{c}{Intensity Reg.} \\
Aug. & Aug.
& Acc$_\text{macro}$ $\uparrow$ & EMR $\uparrow$
& Top-1 Acc. $\uparrow$ & Top-5 Acc. $\uparrow$
& Acc$_\text{num}$ $\uparrow$
& MAE$_\text{mean}$ $\downarrow$ & MAE$_\text{overall}$ $\downarrow$ \\
\midrule

\multirow{2}{*}{None} & None & 91.59 / 82.81 & 58.96 / 30.86 & 21.52 / 5.76 & 47.59 / 18.01 & 61.11 / 45.81 & 0.14 / 0.22 & 0.16 / 0.14 \\
     & Both & \textbf{95.58} / \textbf{86.15} & \textbf{76.48} / \textbf{39.22} & \textbf{36.78} / \textbf{12.19} & \textbf{75.98} / \textbf{32.97} & \textbf{77.24} / \textbf{47.36} & \textbf{0.10} / \textbf{0.19} & 0.16 / 0.17 \\
\midrule

\multirow{2}{*}{Pre}  & None & 84.31 / 78.06 & 39.94 / 22.56 & 11.11 / 3.80 & 27.73 / 11.76 & 53.07 / 41.65 & 0.20 / 0.26 & 0.16 / 0.14 \\
     & Both & \textbf{91.58} / \textbf{83.11} & \textbf{60.71} / \textbf{32.38} & \textbf{23.19} / \textbf{8.81} & \textbf{55.11} / \textbf{25.50} & \textbf{65.53} / \textbf{42.79} & \textbf{0.13} / \textbf{0.21} & 0.16 / 0.17 \\
\midrule

\multirow{2}{*}{Post} & None & 78.29 / 72.64 & 28.45 / 16.64 &  8.01 / 2.56 & 19.95 /  8.07 & 41.12 / 41.18 & 0.24 / 0.30 & 0.18 / 0.15 \\
     & Both & \textbf{90.25} / \textbf{82.38} & \textbf{56.05} / \textbf{30.96} & \textbf{18.06} / \textbf{7.01} & \textbf{44.54} / \textbf{21.08} & \textbf{60.74} / \textbf{42.64} & \textbf{0.15} / \textbf{0.22} & \textbf{0.16} / 0.16 \\
\midrule

\multirow{2}{*}{Either} & None & 81.31 / 75.43 & 34.37 / 19.84 &  9.49 / 3.11 & 23.78 /  9.97 & 47.23 / 41.57 & 0.22 / 0.28 & 0.17 / 0.15 \\
       & Both & \textbf{90.95} / \textbf{82.77} & \textbf{58.57} / \textbf{31.75} & \textbf{20.58} / \textbf{7.82} & \textbf{49.88} / \textbf{23.20} & \textbf{63.23} / \textbf{42.63} & \textbf{0.14} / \textbf{0.22} & \textbf{0.16} / 0.16 \\
\midrule

\multirow{2}{*}{Both} & None & 75.42 / 71.13 & 21.68 / 13.85 &  4.54 / 1.76 & 12.84 /  5.83 & 40.72 / 39.85 & 0.27 / 0.31 & 0.17 / 0.15 \\
     & Both & \textbf{88.48} / \textbf{80.87} & \textbf{49.77} / \textbf{27.58} & \textbf{12.57} / \textbf{5.48} & \textbf{35.20} / \textbf{17.47} & \textbf{56.57} / \textbf{39.78} & \textbf{0.17} / \textbf{0.23} & \textbf{0.16} / 0.16 \\

\bottomrule
\end{tabular}

\end{table*}

\section{Baseline Model}
\label{sec:baseline_model}

\subsection{AudioMAE-Fx: fine-tuning AudioMAE for speech AEI}
\label{subsec:baseline_audiomae}

We propose \textbf{AudioMAE-Fx}, a strong baseline that fine-tunes AudioMAE\,\cite{HuangEtAl22_AudioMAE_NeurIPS} for speech AEI on VoxEffects.
Given an input waveform, we extract log-mel filterbank features and feed them to the AudioMAE backbone.

AudioMAE-Fx is trained in a multi-task manner with lightweight prediction heads on top of the shared AudioMAE representation:
(i) \textbf{Presence} as $K$-way multi-label classification ($K{=}6$ effects),
(ii) \textbf{Preset} as $C$-way classification (here $C{=}2520$),
(iii) \textbf{\#Act} as counting via classification over $\{0,\dots,K\}$,
(iv) \textbf{Intensity (scalar)} via regression, and
(v) \textbf{Intensity (vector)} via $K$-dimensional regression.
All heads are computed from a single forward pass and optimized jointly.

Let $\mathcal{L}_{\text{pres}}$ denote the multi-label presence loss (binary cross-entropy with logits),
$\mathcal{L}_{\text{preset}}$ the preset cross-entropy loss,
$\mathcal{L}_{\#\text{act}}$ the active-count loss, and
$\mathcal{L}_{s}$ / $\mathcal{L}_{v}$ the L1 losses for scalar and vector intensity, respectively.
The overall objective is
$\mathcal{L}=\lambda_{\text{pres}}\mathcal{L}_{\text{pres}}+
\lambda_{\text{preset}}\mathcal{L}_{\text{preset}}+
\lambda_{\#\text{act}}\mathcal{L}_{\#\text{act}}+
\lambda_{s}\mathcal{L}_{s}+
\lambda_{v}\mathcal{L}_{v}$.

\subsection{Training configurations}
\label{subsec:train_configs}

We train AudioMAE-Fx under two configurations that mirror the robustness settings in VoxEffects (\S~\ref{subsec:dist_degrad}).

\smallskip\noindent\textbf{Stage 1: baseline fine-tuning.}
We fine-tune AudioMAE-Fx from a pretrained model trained on AudioSet\,\cite{GemmekeEtAl17_AudioSet_ICASSP} on VoxEffects without applying degradations, i.e., $\mathcal{D}_{\text{pre}}$ and $\mathcal{D}_{\text{post}}$ are set to identity.
All audio is resampled to 16\,kHz for training and evaluation.
We set $\lambda_{\text{pres}}=5$ and all other loss weights to 1.
Training proceeds until validation performance plateaus.
All trainable parameters are optimized using AdamW~\cite{loshchilov2018decoupled} with a base learning rate of $10^{-3}$ and weight decay $0.05$. 
We adopt layer-wise learning-rate decay following~\cite{BaoDongPiaoWei22_BEiT_ICLR}, using a decay factor of $0.75$ across transformer layers. 
Batch size set to $64$.

\smallskip\noindent\textbf{Stage 2: robustness fine-tuning.}
Starting from the Stage-1 checkpoint, we further fine-tune with \textbf{Both} capture- and platform-side degradations.
Specifically, each $\mathcal{D}$ draws two random degradations (additive noise, resampling, quantization, and lossy codecs) and applies them sequentially, applying two degradations before rendering and two after rendering.
This curriculum-learning-style procedure\,\cite{BengioLCW09_CurriculumLearning_ICML} first learns stable AEI cues on clean rendered audio, then adapts to distribution artifacts.
We fine-tune for an additional $50{,}000$ steps.

\section{Evaluation and Discussion}
\label{sec:experiments}

\subsection{Experimental setup}
\label{subsec:exp_setup}

We evaluate AudioMAE-Fx on (i) ID test splits of DAPS/EARS/TSP and (ii) an OOD VCTK test set, reporting results under the five degradation settings in \S~\ref{subsec:dist_degrad}.
Because evaluating all utterances under all 2520 preset tuples is expensive, we report results on fixed subsets: 60 ID utterances (20 per corpus) and 60 OOD utterances from VCTK.

\subsection{Benchmarking results}
\label{subsec:results_main}

Table~\ref{tab:robustness_idood_2col} presents the benchmarking results.
We compare two training configurations introduced in \S~\ref{subsec:train_configs}: a baseline model trained without augmentation and a model further fine-tuned with degradations.
Robustness fine-tuning consistently improves performance across all test conditions, improving presence detection on both ID and OOD and substantially boosting Top-1/Top-5 accuracy of the preset classification. 
Note that preset classification is challenging (2520 perceptually overlapping classes), leading to low Top-1 accuracy.

More importantly, under mismatched or stronger test degradations, the baseline without robustness fine-tuning degrades sharply, especially on OOD, whereas the model fine-tuned with robustness augmentation maintains consistently higher accuracy, indicating improved invariance to degradation artifacts and domain shift. 
For intensity regression, robustness fine-tuning also reduces MAE$_\text{mean}$ across settings, while MAE$_\text{overall}$ remains broadly comparable, which indicates an improvement space in future work.
Overall, incorporating degradations during training is critical for stable multi-label presence detection and auxiliary predictions when test conditions differ from training.

We next analyze where performance differs across effects via effect-wise evaluation, and how much acoustic context is needed via duration-controlled testing.

\subsection{Effect-wise analysis}
\label{subsec:effectwise}

\begin{figure}[h]
    \centering
    \includegraphics[width=\columnwidth]{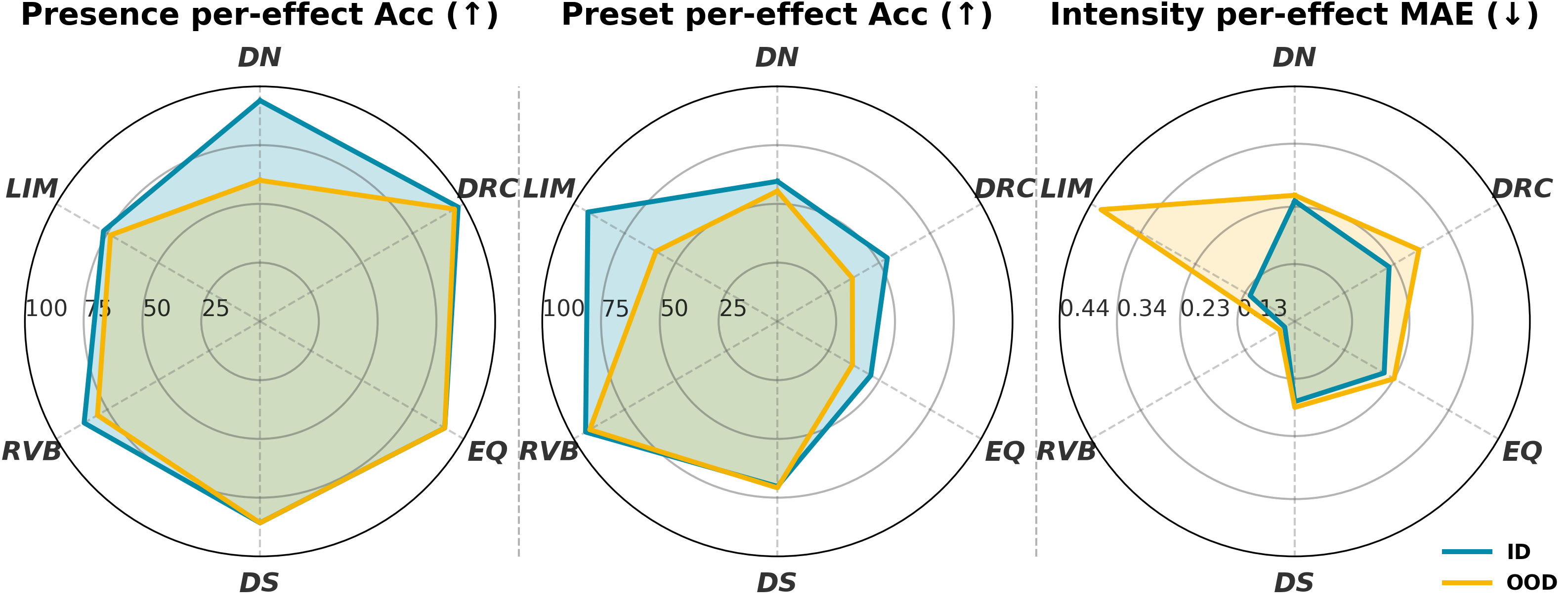}
    \caption{Effect-wise radar analysis. Each panel reports evaluation metrics on In-Domain vs. Out-of-Domain test.}
    \label{fig:effectwise_radar}
\end{figure}

We analyze effect-wise performance across the six effects.
Fig.~\ref{fig:effectwise_radar} shows per-effect metrics for the model fine-tuned with degradation augmentation, including presence accuracy, preset Top-1 accuracy, and intensity MAE.

The effect-wise results show clear heterogeneity in cross-domain generalization.
For presence detection, most effects exhibit only small ID$\rightarrow$OOD gaps, whereas \textbf{\textit{DN}} degrades substantially, likely because corpus-dependent recording characteristics (even in clean speech) affect how DN presets modify low-energy regions, thereby altering the DN-induced artifacts used for detection.
For fine-grained preset classification, \textbf{\textit{RVB}} transfers relatively well, while \textbf{\textit{DRC}} and \textbf{\textit{LIM}} show larger OOD drops, indicating higher sensitivity of preset recognition to corpus-dependent dynamics and loudness characteristics. For intensity regression, the errors are broadly similar across domains except for \textbf{\textit{LIM}}. This is consistent with limiter behavior depending strongly on level distributions and peak statistics that vary across recording conditions.
Overall, these patterns motivate effect-specific strategies, such as targeted augmentation and domain balancing for the most fragile effects.

\subsection{Input duration analysis}
\label{subsec:duration}

We next analyze the impact of input duration on AEI task, using OOD test and the model with degradation augmentation.
We crop waveforms to \mbox{\{0.2, 0.5, 1, 2, 3, 4, 5\}} seconds and report per-effect presence accuracy together with Acc$_\text{macro}$ and EMR. 

\begin{figure}[h]
    \centering
    \includegraphics[width=\linewidth]{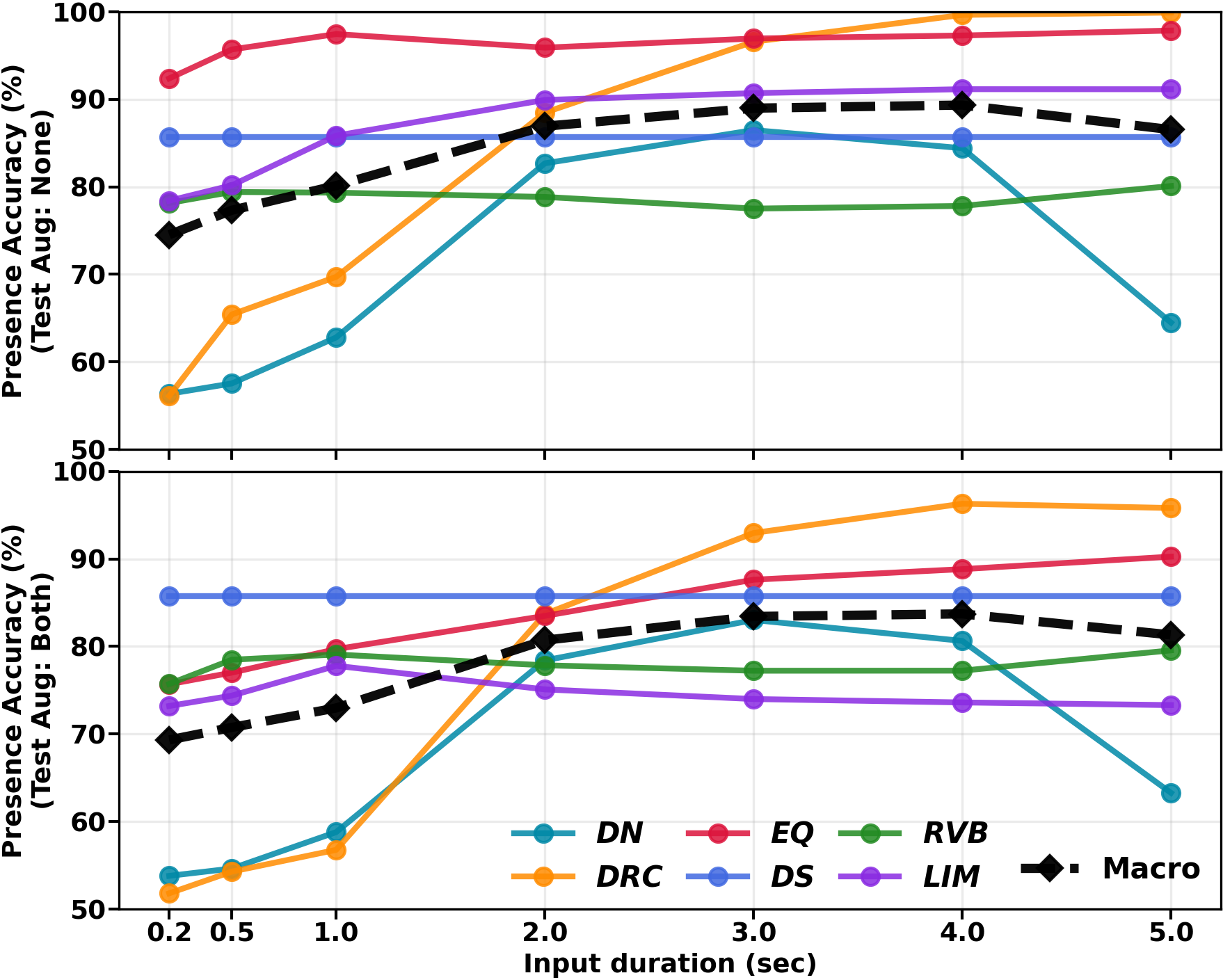}
    \caption{Presence detection accuracy with various input duration from 0.2\,s to 5\,s on OOD test.}
    \label{fig:duration_analysis}
\end{figure}

Fig.~\ref{fig:duration_analysis} shows that longer inputs generally improve presence detection: Acc$_\text{macro}$ and EMR increase markedly, indicating that reliable multi-label decisions require several seconds of acoustic context.
Under test-time degradations, performance drops and the trend becomes less monotonic, suggesting that degradation artifacts can partially dominate the decision boundary and reduce the marginal benefit of additional context.

Effect-wise trends highlight different cue requirements.
\textbf{\textit{DN}} benefits from non-speech context (noise-floor characteristics), so very short crops are insufficient and longer crops can dilute the cues as speech dominates.
In contrast, \textbf{\textit{RVB}} and \textbf{\textit{DS}} remain relatively stable with shorter inputs because their signatures are locally observable.
\textbf{\textit{LIM}} is more sensitive to domain shift and audio duration since activation depends on level and peak statistics, while \textbf{\textit{DRC}} typically benefits from longer segments that contain richer level variation and envelope dynamics. 

Overall, duration analysis clarifies how much temporal context each effect requires and helps guide practical design for AEI systems, such as selecting input window lengths and prioritizing duration-aware training or effect-specific augmentation.

\subsection{Gender fairness analysis}
\label{subsec:gender_fairness}

\begin{figure}[htbp]
    \centering
    \includegraphics[width=\columnwidth]{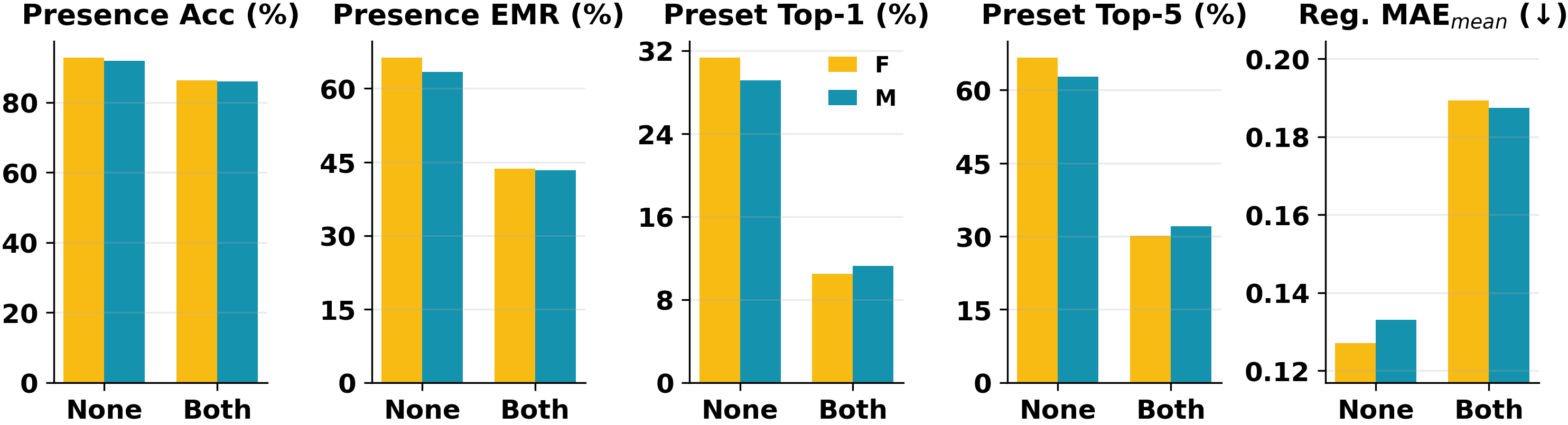}
    \caption{Gender fairness visualization.}
    \label{fig:gender_analysis}
\end{figure}

Across the combined training set, gender is broadly balanced (female 50.5\%, male 40.8\%), while 8.7\% of unknown labels. We analyze gender fairness by evaluating the model on two subsets of 60 utterances each, rendered with the same effect chain.
Fig.~\ref{fig:gender_analysis} shows closely matched performance under both \textit{None} and \textit{Both} degradation settings.
The primary performance drop is driven by degradations rather than gender.
A larger audit with content-controlled matching is left for future work.

\section{Limitations}
\label{sec:limitations}

VoxEffects has several limitations.
First, it assumes a fixed post-production chain and a finite preset bank, which yields tractable supervision but does not cover alternative orderings, repeated stages, or continuously tuned parameters found in real workflows.
Second, rendering relies on a single effect implementation stack (\texttt{Pedalboard}\,\cite{sobot_pedalboard_2023}); models may therefore face cross-implementation mismatch when deployed on audio processed by other commercial toolchains, motivating multi-engine and multi-variant renderers in future releases.
Third, difficulty is effect-dependent: cues can be subtle for conservative denoising and limiter behavior depends strongly on upstream level/dynamics, which weakens the alignment between preset labels and observable artifacts, especially for fine-grained preset classification and intensity prediction under domain shift.

\section{Conclusion}
\label{sec:conclusion}

We introduced VoxEffects, a speech-oriented dataset and benchmark for AEI, where the goal is to infer which post-production effects are present in a processed waveform and how strongly they are applied. VoxEffects is built from clean speech recordings and a quality-oriented speech post-production chain with curated preset banks, providing exact supervision at multiple granularities. We also proposed a robustness protocol that applies controlled capture-side and platform-side degradations to assess generalization under realistic distribution artifacts.

As a strong baseline, we presented \textbf{AudioMAE-Fx}, which fine-tunes AudioMAE with multi-task heads and audio degradation. Our experiments show that speech AEI is feasible but sensitive to domain shift and distribution degradations, and that robustness-oriented training with degradations improves performance across test-time conditions. Future work includes scaling evaluation on more diverse real-world post-produced speech, expanding the effect vocabulary and preset regimes, and extending the benchmark to settings with time-varying processing.

\section{Acknowledgments}
This study is supported by a project (JPNP22007) commissioned by the New Energy and Industrial Technology Development Organization (NEDO). This study was carried out using the TSUBAME4.0 supercomputer at Institute of Science Tokyo.


\bibliographystyle{IEEEtran}
\bibliography{mybib}

\section{Appendix}
Table~\ref{tab:robustness_idood_5x5_appendix} reports a comprehensive robustness evaluation under a full train/test degradation grid.
We consider five augmentation settings applied at training and/or evaluation time: \textit{None}, \textit{Pre} (pre-effect), \textit{Post} (post-effect), \textit{Either} (randomly pre or post), and \textit{Both} (pre and post).
Each cell is written as \textit{In-Domain / Out-of-Domain}, where In-Domain aggregates results on the training-domain corpora (DAPS/EARS/TSP) and Out-of-Domain corresponds to VCTK.
We summarize performance on four tasks: effect presence detection (macro accuracy and exact-match ratio), preset identification (Top-1/Top-5), number-of-active-effects classification (Acc$_\text{num}$), and intensity estimation (mean MAE and overall MAE; lower is better).
\begin{table*}[!htbp]
\centering
\caption{Appendix: Benchmarking with full train/test degradation grid. Entries are In-Domain / Out-of-Domain.}
\label{tab:robustness_idood_5x5_appendix}
\small
\begin{tabular}{llccccccc}
\toprule
\multicolumn{1}{c}{Test} & \multicolumn{1}{c}{Train}
& \multicolumn{2}{c}{Presence}
& \multicolumn{2}{c}{Preset}
& \multicolumn{1}{c}{\#Active}
& \multicolumn{2}{c}{Intensity} \\
aug. & aug.
& Acc$_\text{macro}$ $\uparrow$ & EMR $\uparrow$
& Top-1 $\uparrow$ & Top-5 $\uparrow$
& Acc$_\text{num}$ $\uparrow$
& MAE$_\text{mean}$ $\downarrow$ & MAE$_\text{overall}$ $\downarrow$ \\
\midrule

None  & None   & 91.59 / 82.81 & 58.96 / 30.86 & 21.52 / 5.76 & 47.59 / 18.01 & 61.11 / 45.81 & 0.14 / 0.22 & 0.16 / 0.14 \\
None  & Pre    & 95.73 / 86.63 & 77.04 / 41.41 & 33.85 / 12.90 & 74.42 / 34.76 & 77.98 / 53.60 & 0.09 / 0.17 & 0.16 / 0.16 \\
None  & Post   & 95.57 / 85.37 & 76.36 / 36.88 & 31.57 / 10.20 & 71.34 / 28.33 & 77.11 / 46.75 & 0.10 / 0.20 & 0.16 / 0.17 \\
None  & Either & 95.66 / 86.06 & 76.74 / 39.50 & 35.62 / 13.27 & 73.53 / 34.54 & 77.80 / 48.86 & 0.10 / 0.19 & 0.16 / 0.17 \\
None  & Both   & 95.58 / 86.15 & 76.48 / 39.22 & 36.78 / 12.19 & 75.98 / 32.97 & 77.24 / 47.36 & 0.10 / 0.19 & 0.16 / 0.17 \\
\midrule

Pre   & None   & 84.31 / 78.06 & 39.94 / 22.56 & 11.11 / 3.80 & 27.73 / 11.76 & 53.07 / 41.65 & 0.20 / 0.26 & 0.16 / 0.14 \\
Pre   & Pre    & 92.80 / 83.96 & 64.87 / 34.61 & 26.80 / 9.77 & 58.61 / 28.10 & 68.96 / 47.56 & 0.11 / 0.19 & 0.16 / 0.16 \\
Pre   & Post   & 91.73 / 82.52 & 61.13 / 30.62 & 20.57 / 7.70 & 45.26 / 22.70 & 64.89 / 41.92 & 0.15 / 0.22 & 0.17 / 0.18 \\
Pre   & Either & 91.87 / 83.21 & 62.08 / 32.71 & 23.47 / 9.89 & 54.11 / 27.23 & 66.67 / 43.38 & 0.16 / 0.21 & 0.16 / 0.17 \\
Pre   & Both   & 91.58 / 83.11 & 60.71 / 32.38 & 23.19 / 8.81 & 55.11 / 25.50 & 65.53 / 42.79 & 0.13 / 0.21 & 0.16 / 0.17 \\
\midrule

Post  & None   & 78.29 / 72.64 & 28.45 / 16.64 & 8.01 / 2.56 & 19.95 / 8.07 & 41.12 / 41.18 & 0.24 / 0.30 & 0.18 / 0.15 \\
Post  & Pre    & 88.38 / 80.43 & 49.53 / 28.46 & 14.30 / 6.74 & 42.87 / 19.79 & 54.76 / 46.51 & 0.16 / 0.23 & 0.17 / 0.15 \\
Post  & Post   & 89.19 / 82.46 & 52.58 / 30.94 & 19.13 / 8.05 & 47.48 / 22.93 & 61.31 / 42.39 & 0.15 / 0.22 & 0.16 / 0.16 \\
Post  & Either & 89.46 / 82.46 & 51.15 / 30.78 & 19.71 / 8.02 & 47.81 / 22.81 & 61.61 / 42.39 & 0.15 / 0.22 & 0.16 / 0.16 \\
Post  & Both   & 90.25 / 82.38 & 56.05 / 30.96 & 18.06 / 7.01 & 44.54 / 21.08 & 60.74 / 42.64 & 0.15 / 0.22 & 0.16 / 0.16 \\
\midrule

Either & None   & 81.31 / 75.43 & 34.37 / 19.84 & 9.49 / 3.11 & 23.78 / 9.97 & 47.23 / 41.57 & 0.22 / 0.28 & 0.17 / 0.15 \\
Either & Pre    & 90.61 / 82.25 & 57.47 / 31.72 & 17.16 / 8.26 & 49.18 / 24.01 & 61.58 / 46.94 & 0.13 / 0.21 & 0.16 / 0.16 \\
Either & Post   & 91.38 / 82.52 & 60.34 / 30.74 & 17.66 / 7.42 & 50.13 / 22.20 & 64.75 / 41.79 & 0.13 / 0.22 & 0.16 / 0.17 \\
Either & Either & 91.39 / 82.87 & 59.98 / 31.99 & 21.95 / 8.96 & 51.27 / 25.10 & 65.99 / 42.83 & 0.14 / 0.22 & 0.16 / 0.16 \\
Either & Both   & 90.95 / 82.77 & 58.57 / 31.75 & 20.58 / 7.82 & 49.88 / 23.20 & 63.23 / 42.63 & 0.14 / 0.22 & 0.16 / 0.16 \\
\midrule

Both  & None   & 75.42 / 71.13 & 21.68 / 13.85 & 4.54 / 1.76 & 12.84 / 5.83 & 40.72 / 39.85 & 0.27 / 0.31 & 0.17 / 0.15 \\
Both  & Pre    & 88.63 / 79.78 & 45.27 / 26.34 & 13.71 / 5.73 & 37.45 / 17.80 & 52.76 / 43.33 & 0.17 / 0.23 & 0.17 / 0.16 \\
Both  & Post   & 90.21 / 80.75 & 52.30 / 27.06 & 14.17 / 5.66 & 45.02 / 17.92 & 57.05 / 38.74 & 0.17 / 0.23 & 0.16 / 0.17 \\
Both  & Either & 90.25 / 80.94 & 57.78 / 27.67 & 14.10 / 6.25 & 41.25 / 19.04 & 55.80 / 39.32 & 0.17 / 0.23 & 0.16 / 0.16 \\
Both  & Both   & 88.48 / 80.87 & 49.77 / 27.58 & 12.57 / 5.48 & 35.20 / 17.47 & 56.57 / 39.78 & 0.17 / 0.23 & 0.16 / 0.16 \\
\bottomrule
\end{tabular}
\end{table*}

\end{document}